\documentclass[final,5p,times,twocolumn]{elsarticle}
\biboptions{sort&compress}

\usepackage{epsfig}

\usepackage{amssymb}
\usepackage{amsmath}
\usepackage{lipsum}
\usepackage{amsthm}
\usepackage{hyperref}

\journal{Physics Letters B}

\begin{document}

\begin{frontmatter}




\author[inst1]{A. Anupam}
\author[inst2]{C. Dash}
\author[inst3]{Yashraj}
\author[inst3]{R. P. Singh}
\author[inst1]{B. B. Sahu\corref{cor1}}
\cortext[cor1]{Corresponding author}
\ead{bbsahufpy@kiit.ac.in}

\affiliation[inst1]{organization={Department of Physics, School of Applied Sciences, KIIT Deemed to be University},
addressline={},
city={Bhubaneswar},
postcode={751024},
state={Odisha},
country={India}}

\affiliation[inst2]{organization={Department of Physics, Anchalika Mahavidyalaya Gadia},
addressline={},
city={Mayurbhanj},
postcode={757023},
state={Odisha},
country={India}}

\affiliation[inst3]{organization={Inter University Accelerator Centre},
addressline={Aruna Asaf Ali Marg},
city={New Delhi},
postcode={110067},
state={Delhi},
country={India}}

\title{$N = 32$ Shell Evolution in Calcium Isotopes: Evidence from Macroscopic to Microscopic Models}

\begin{abstract}

The well-established N=32 subshell closure in calcium serves as an important benchmark for nuclear structure evolution toward neutron-rich nuclei. Recent charge-radius measurements of potassium isotopes [Nat. Phys. 17, 439-443 (2021)] reveal no corresponding signature of enhanced stability at the same neutron number. Motivated by this discrepancy, we investigate the structural evolution of even-even Ca isotopes from N=22 to 36 using macroscopic, relativistic mean-field (RMF), and large-scale shell-model approaches, with the KB3G, GXPF1A, GXPF1B, and FPD6 interactions in the KSHELL framework. The macroscopic analysis, evaluated against seven theoretical frameworks and NNDC data, reveals a coherent change in neutron structure at N=32, most clearly in the shell-gap energy $\Delta E$ and the three-point binding-energy filter, corroborated by the charge-radius filter. The RMF spectrum with NL3* shows a clear $\nu2p_{3/2}$-$\nu2p_{1/2}$ separation consistent with this signature. KB3G and FPD6 reproduce the N=32 enhancement of $E(2_1^+)$ and reduction of $R_{4/2}$, whereas GXPF1A and GXPF1B give a weaker N=32 signature than N=34 signature, the reverse of the experimental pattern, showing that the two subshells depend differently on the monopole interaction. The deformation parameter and B(E2) values have been obtained with above four interactions and are in good agreement with the experimental results.

\end{abstract}




\end{frontmatter}




\section{Introduction}
\label{sec:intro}

The atomic nucleus, a complex system of protons and neutrons bound together by the strong nuclear force, exhibits a rich variety of structural phenomena. Understanding the intricate details of nuclear structure remains a central goal of nuclear physics. A cornerstone of our understanding is the nuclear shell model, which, akin to the electronic shell model in atoms, describes how nucleons arrange themselves into quantized energy levels, forming shell structures \cite{mayer1949closed, honma2006structure, wienholtz2013masses, steppenbeck2013evidence, garcia2016unexpectedly}. These shells arise from the interplay between the average potential experienced by each nucleon and the residual interactions between them. Nuclei exhibiting closed-shell configurations at specific proton or neutron numbers, commonly referred to as magic numbers (2, 8, 20, 28, 50, 82, and 126), demonstrate markedly greater binding energies and structural resilience relative to adjacent isotopes. This enhanced stability arises from substantial energy gaps between major shells, resulting in minimal configuration mixing and reduced level density near the Fermi surface.

However, recent investigations of exotic nuclei far from the line of $\beta$ stability have revealed that these magic numbers are not immutable \cite{otsuka2001magic, janssens2002structure, liddick2004development, dinca2005reduced, perrot2006beta, sorlin2008nuclear, coraggio2014realistic, michimasa2018magic, doornenbal2018spectroscopy, rodriguez2020robustness, leistenschneider2021precision, an2024improved}. As the neutron-to-proton ratio deviates substantially from unity, the nuclear landscape undergoes significant changes, with some traditional magic numbers weakening or disappearing while new ones emerge. This evolution of shell structure, commonly referred to as ``shell evolution,'' is driven by the interplay of the tensor component of the nucleon--nucleon interaction, modifications of spin--orbit splittings, and other components of the effective nuclear interaction \cite{otsuka2010three}. The exploration of neutron-rich nuclei therefore remains at the forefront of contemporary nuclear structure research, propelled by rapid advances in rare-isotope beam facilities and by the development of sophisticated theoretical frameworks, including large-scale shell-model calculations, \textit{ab initio} methods, and energy-density functional approaches.

Among neutron-rich systems, Calcium (Ca) isotopes with their proton number $Z = 20$ have attracted considerable attention due to their role as benchmark systems for the testing of nuclear shell evolution \cite{wienholtz2013masses, steppenbeck2013evidence, garcia2016unexpectedly, coraggio2014realistic, michimasa2018magic, hagen2016neutron}. Lying along the $N = Z$ line up to $^{40}$Ca, the calcium isotopes extend far into the neutron-rich region, providing access to a wide range of neutron numbers while maintaining a closed proton shell that simplifies theoretical interpretation of neutron shell effects. While the conventional magic numbers $N = 8$, 20, and 28 are well established in Ca isotopes, recent experimental and theoretical studies have revealed pronounced subshell closures at $N = 32$ \cite{hagen2012evolution, crawford2010beta, garcia2015ground, bhoy2020shell, akkoyun2020nuclear, enciu2022extended, li2024spectroscopy, liu2026shell} and $N = 34$ \cite{steppenbeck2013evidence, garcia2015ground, mcgrory1970shell, rejmund2007shell, coraggio2009spectroscopic}, underscoring the need for a deeper understanding of the underlying nuclear forces in this region of the nuclear chart. The emergence of the $N = 32$ sub-shell closure in Ca isotopes has been experimentally investigated since the early measurement of the first $2^+$ excitation energy at ISOLDE in 1985 \cite{huck1985beta, miller2019proton}, marking the beginning of sustained experimental and theoretical efforts to characterize this sub-shell closure. The magic structures and shell evolution of Ca isotopes have since attracted intensive investigation. The high excitation energy of the first $2^+$ state in $^{54}$Ca, measured at RIKEN by Steppenbeck et al. (2013) \cite{steppenbeck2013evidence}, provided definitive evidence for a sizable sub-shell closure at $N = 34$, prompting the development of improved effective interactions such as GXPF1B to reproduce this strong closure.

Despite extensive investigation of calcium isotopes, a contemporary puzzle has emerged that motivates the present comprehensive study. Recent precision charge-radius measurements of exotic potassium isotopes by Á. Koszorús et al. (2021) \cite{koszorus2021charge} revealed no pronounced signature of magicity at $N = 32$ in the evolution of the charge radii, in contrast to the enhanced shell effects inferred from spectroscopic studies of calcium isotopes at the same neutron number.  This apparent contradiction raises an important question: Does $N=32$ shell closure in calcium arise naturally from the underlying nuclear interaction and manifest differently in potassium because of the removal of a proton from the $f_{7/2}$ orbital ($Z = 19$ versus $Z = 20$), or does it require exotic mechanisms beyond conventional theoretical descriptions? Addressing this question requires a systematic investigation using complementary theoretical approaches to determine whether established shell-model interactions can consistently reproduce the observed signatures of the $N=32$ shell structure in calcium and account for its evolution across neighbouring isotopic and isotonic chains.

To address this question comprehensively, we present a systematic study of the nuclear structure of $^{42-56}$Ca isotopes using a three-pronged investigation combining model-independent macroscopic analysis, single-particle structure examination through relativistic mean field (RMF) calculations, and detailed microscopic shell model validation. Our first approach employs macroscopic observables \cite{koszorus2021charge, swain2019structure, wang2013shell}: two-neutron separation energies $(S_{2n})$, three-point binding energy filters $(\Delta_{1n}^{(3)} B.E.)$, shell gap energies $(\Delta E)$, and charge radius evolution calculated for experimental data and compared systematically against predictions from seven independent theoretical frameworks: RCHB \cite{xia2018limits}, FRDM \cite{moller2016nuclear}, dHRB \cite{guo2024nuclear}, and four RHB parametrizations (PC-L3R, PC-X, DD-MEX, DD-PCX) \cite{liu2023nuclear}. These macroscopic indicators, not subject to shell model truncation limitations, provide model-independent evidence for shell structure evolution. Second, we examine the single-particle orbital structure through RMF calculations \cite{dash2024examining} to identify the fundamental orbital energy gaps underlying observed macroscopic signatures. Third, we perform large-scale shell model calculations using the KSHELL code \cite{shimizu2013nuclear,shimizu2019thick} which implements a highly efficient M-scheme shell model diagonalization framework, in which the many-body wave functions are constructed as linear combinations of Slater determinants, each representing a specific configuration of nucleons in the available single-particle orbits. This approach allows for a complete and unbiased treatment of the many-body correlations between valence nucleons. However, the M-scheme basis leads to very large-dimensional matrices, requiring efficient computational techniques. The thick-restart block Lanczos method is used to solve the large-scale eigenvalue problem. This iterative method efficiently extracts the lowest eigenvalues and eigenvectors of a sparse, symmetric matrix, such as the shell model Hamiltonian. The block Lanczos method accelerates the convergence by simultaneously treating a block of vectors, while the thick-restart technique reduces the computational cost by periodically restarting the iteration with a refined set of Lanczos vectors \cite{shimizu2019thick}.

We employ the KB3G interaction \cite{abzouzi1991influence, poves2001shell,bally2019variational}, a modified version of the Kuo-Brown G-matrix interaction derived from realistic nucleon-nucleon potentials, alongside GXPF1A \cite{honma2002effective, honma2005shell}, developed through extensive least-squares fitting to experimental spectra across the pf-shell region. Additionally, we include GXPF1B \cite{steppenbeck2013evidence,honma2008shell}, refined with monopole corrections specifically optimized for the strong $N = 34$ closure, and FPD6 \cite{richter1991new}, an alternative pf-shell parametrization. Calculations are performed within the full pf-shell model space $(1f_{7/2},\,2p_{3/2},\,1f_{5/2},\,2p_{1/2})$ orbitals with $^{40}$Ca as the inert core, employing the thick-restart block Lanczos method for efficient diagonalization of large-dimensional Hamiltonian matrices. Emphasis is placed on the evolution of collectivity as inferred from excitation energies, $R_{4/2}$ energy ratios, reduced electromagnetic transition probabilities B(E2), and intrinsic quadrupole deformation parameters $\beta_2$. The reliability of employed effective interactions is critically assessed through systematic comparison with available experimental data and validation against our independent macroscopic analysis results. This comprehensive cross-validation between complementary approaches yields substantially stronger conclusions than any single method could offer independently.

This investigation aims to provide comprehensive understanding of $N = 32$ subshell closure physics in calcium isotopes through multi-faceted validation, establishing whether standard pf-shell interactions capture the observed signatures without exotic modifications and thereby elucidating the physical origin of the apparent calcium-potassium discrepancy. By comparing results obtained with four interactions constructed through fundamentally different philosophies, we assess the robustness of theoretical predictions and identify where standard frameworks succeed versus where they encounter fundamental limitations requiring future developments.

The structure of this paper is as follows: Sec.~\ref{subsec:ground-state} presents the theoretical framework for macroscopic analysis, including detailed formulations of all observables and the theoretical frameworks employed for systematic comparison. Sec.~\ref{subsec:SPS} describes the relativistic mean field approach for examining single-particle orbital structure in a very brief manner. Sec.~\ref{subsec:LSSM} outlines the shell model Hamiltonian formulation, the KSHELL computational methodology, and the four effective interactions employed in microscopic calculations. Sec.~\ref{subsec:macroscopic} presents comprehensive results from macroscopic analysis, revealing $N = 32$ signatures across multiple observables and theoretical frameworks and Sect.~\ref{subsec:SPE-RMF} RMF predictions of orbital energy gaps underlying the observed macroscopic signatures. Sect.~\ref{subsec:micro-shell-model} analyzes microscopic shell model results, comparing four interactions' predictions for spectroscopic and electromagnetic observables. Also, Sec.~\ref{sec:results} provides integrated discussion synthesizing results from all three approaches, establishing the convergence between model-independent macroscopic evidence, single-particle orbital physics, and microscopic shell model predictions. Finally, Sec.~\ref{sec:conclusions} presents our conclusions regarding $N = 32$ subshell closure physics and implications for understanding shell evolution in neutron-rich isotopes.

\section{Theoretical framework}
\label{sec:theoretical-framework}
The present investigation employs three complementary theoretical approaches to characterize the structural evolution of even-mass Ca isotopes spanning $N = 22-36$. The macroscopic framework provides model-independent ground-state indicators of shell closure through bulk nuclear observables evaluated against an ensemble of theoretical mass models and energy density functionals, establishing thermodynamic fingerprints of shell structure evolution that are free of valence-space truncation errors. The relativistic mean-field framework affords direct access to the single-particle level structure and spin-orbit splitting underlying the macroscopic signatures. The large-scale shell model framework provides spectroscopic observables and effective single-particle energies that establish the quantum mechanical origin of the $N = 32$ subshell closure. The theoretical foundations of each approach are described in the subsections below.

\subsection{Model Independent Macroscopic Systematics: Ground-State Observables} \label{subsec:ground-state}
Shell and subshell closures manifest in systematic trends of bulk nuclear ground-state properties as characteristic discontinuities superimposed on otherwise smooth isotopic evolutions. In the present work, six complementary observables - derived from nuclear binding energies and charge radii, are evaluated across the Ca-isotopic chain, following the formalisms of R. R. Swain and B. B. Sahu \cite{swain2019structure} and Á. Koszorús et al.\cite{koszorus2021charge}. These quantities are computed from experimental data and systematically compared with predictions from seven theoretical frameworks, enabling identification of robust signatures of the closure of $N = 32$ subshell closure across multiple and methodologically distinct indicators.

The two-neutron separation energy, defined as,
\begin{equation}
S_{2n} (N, Z) = B.E.(N,Z) - B.E.(N-2,Z)
\label{eq:S2n}
\end{equation}
quantifies the binding energy contribution of the outermost neutron pair. At shell closures, $S_{2n}$  exhibits a pronounced downward discontinuity reflecting the abrupt change in binding associated with the opening of a shell gap at the Fermi surface. 

The differential two-neutron separation energy,
\begin{equation}
dS_{2n} (N, Z)
=
\frac{S_{2n}(N+2,Z)-S_{2n}(N,Z)}{2}
\label{eq:dS2n}
\end{equation}
amplifies the sensitivity to shell gap evolution by extracting the local curvature of the $S_{2n}$ surface. This observable exhibit pronounced minima at magic neutron numbers where the binding energy undergoes an abrupt structural change. 

The three-point mass filter,
\begin{equation}
\begin{aligned}[b]
\Delta_{1n}^{(3)} B.E.(N,Z)
&= \frac{1}{2}(-1)^{N}
\big[ B.E.(N+1,Z) - 2B.E.(N,Z) \\
&\qquad + B.E.(N-1,Z) \big]
\end{aligned}
\label{eq:three_point_mass_filter}
\end{equation}
isolates odd-even mass staggering and pairing correlations while suppressing the smooth mass-number-dependent background, thereby revealing quantum shell structure effects with enhanced resolution \cite{koszorus2021charge}.

Following the approach proposed by Koszorús et al., the neutron shell gap can be extracted from the difference of these filters through,
\begin{equation}
\Delta E
=
2
\left[
\Delta_{1n}^{(3)}B.E.(N,Z)
-
\Delta_{1n}^{(3)}B.E.(N+1,Z)
\right]
\label{eq:neutron_shell_gap}
\end{equation}
provides an alternative probe of shell effects through energy level spacings.

The nuclear charge radii \& rms charge radii, incorporating shell correction and quadrupole and hexadecapole deformation contributions following Wang and Li \cite{wang2013shell}, is expressed as
\begin{equation}
r_{\mathrm{ch}}
=
\sqrt{\frac{3}{5}}\,R_C
\left[
1+
\frac{5}{8\pi}
\left(
\beta_2^2+\beta_4^2
\right)
\right]
\label{eq:charge_radius}
\end{equation}

where the nuclear charge radius parameter includes shell gap (Eq.~\ref{eq:neutron_shell_gap}) contributions,
\begin{equation}
R_C
=
r_0 A^{1/3}
+
r_1 A^{-2/3}
+
r_s I(1-I)
+
r_d\frac{\Delta E}{A}
\label{eq:RC}
\end{equation}
with $I = (N-Z)/A$, and parameters $r_0 = 1.2260$ fm, $r_1 = 2.86$ fm, $r_s = -1.09$, and
$r_d = 0.99$ MeV$^{-1}$ fm. The quadrupole ($\beta_2$) and hexadecapole
($\beta_4$) deformation parameters are adopted from the FRDM (2012)
tabulation \cite{moller2016nuclear}.

The three-point charge radius filter,
\begin{equation}
\begin{aligned}[b]
\Delta_{1n}^{(3)}r_{\mathrm{ch}}(N,Z)
=
&\frac{1}{2}(-1)^{N+1}
\big[ r_{\mathrm{ch}}(N+1,Z)
-2r_{\mathrm{ch}}(N,Z) \\
&\qquad + r_{\mathrm{ch}}(N-1,Z) \big]
\end{aligned}
\label{eq:three_point_charge_radius}
\end{equation}
represents a sensitive probe of shell structure, amplifying local discontinuities in the charge radius evolution while suppressing the smooth mass-dependent background, thereby revealing subshell closure signatures that may remain obscured in the binding energy observables.

Experimental binding energies are adopted from the NNDC nuclear database \cite{NNDC_NuDat3}, while experimental charge radii are taken from the compilation of Angeli and Marinova \cite{angeli2013table} and the ADNDT (2021) tabulation \cite{li2021compilation}. The six observables defined in Eqs.~\ref{eq:S2n} to \ref{eq:three_point_charge_radius} are evaluated against predictions from seven theoretical frameworks spanning distinct nuclear structure approaches: the Relativistic Continuum Hartree-Bogoliubov model (RCHB, 2018) \cite{xia2018limits}, the Finite-Range Droplet Model (FRDM, 2012) \cite{moller2016nuclear}, the deformed Hartree-Bogoliubov approach (dHRB, 2024) \cite{guo2024nuclear}, and four relativistic Hartree-Bogoliubov parametrizations employing the PC-L3R, PC-X, DD-MEX, and DD-PCX energy density functionals \cite{liu2023nuclear}. This ensemble collectively encompasses relativistic mean-field, non-relativistic Hartree-Fock-Bogoliubov, and macroscopic-microscopic theoretical methodologies. Structural features that emerge consistently across this full ensemble are thereby established as physically robust, model-independent signatures of subshell closure, distinguishable from predictions that are specific to a particular theoretical approximation.

\subsection{Mean-Field Single-Particle Structure} 
\label{subsec:SPS}
The single-particle structure of the calcium isotopes is examined within the relativistic mean-field (RMF) framework \cite{dash2024examining, gambhir1990relativistic, lalazissis1997new, swain2018nuclear}, which has demonstrated considerable success in describing the ground-state properties of nuclei across the periodic table. In this approach, the nucleus is treated as a system of Dirac nucleons whose interactions are mediated by the exchange of classical meson fields rather than by direct nucleon-nucleon forces. The scalar-isoscalar $\sigma$ field provides the intermediate-range attraction, the vector-isoscalar $\omega$ field accounts for the short-range repulsion, and the vector-isovector $\rho$ field introduces the isospin asymmetry. A distinguishing feature of the RMF framework is that the spin-orbit interaction, which is central to the emergence of nuclear magic numbers, arises naturally from the covariant structure of the theory through the interplay of the large scalar and vector potentials, without requiring any additional parametrization. The nucleonic single-particle states are obtained as self-consistent solutions of the coupled Dirac equation for the nucleons and Klein-Gordon equations for the meson fields, solved iteratively under the mean-field approximation. 

In the present work, all RMF calculations are performed using the NL3* parametrization \cite{lalazissis2009effective}, which represents a refined calibration of the widely used NL3 force \cite{lalazissis1997new} and has been shown to yield improved descriptions of nuclear radii and binding energies across medium and heavy nuclei. The calculated neutron single-particle level spectra in the pf shell are examined as a function of neutron number across the calcium chain to identify the mean-field signatures of the $N = 32$ sub-shell gap and to establish a direct connection between the macroscopic observables discussed in Sec.~\ref{subsec:ground-state} and the microscopic shell model analysis of Sec.~\ref{subsec:LSSM}.

\subsection{Large-Scale Shell Model Calculations} \label{subsec:LSSM}
The shell-model Hamiltonian for nuclear systems is formulated using single-particle energies (T) and two-body nucleon interactions (V) as;
\begin{equation}
H=T+V
\label{eq:shell_model_hamiltonian}
\end{equation}

with

\begin{equation}
T=\sum_{\alpha}\epsilon_{\alpha}\hat{n}_{\alpha}
\label{eq:single_particle_hamiltonian}
\end{equation}

\begin{equation}
V=
\sum_{\substack{
\alpha\leq\beta,\,
\gamma\leq\delta,\,J,M
}}
\left\langle
j_{\alpha}j_{\beta}
\middle|V\middle|
j_{\gamma}j_{\delta}
\right\rangle_J
A_{\alpha,\beta}^{\dagger}(J,M)
A_{\gamma,\delta}(J,M)
\label{eq:two_body_interaction}
\end{equation}

where

\begin{equation}
\hat{n}_{\alpha}
=
\sum_{m_{\alpha}}
c_{\alpha,m_{\alpha}}^{\dagger}
c_{\alpha,m_{\alpha}}
\label{eq:number_operator}
\end{equation}

\begin{equation}
A_{\alpha,\beta}^{\dagger}(J,M)
=
\frac{1}{\sqrt{1+\delta_{\alpha\beta}}}
\sum_{m_{\alpha},m_{\beta}}
\left\langle
j_{\alpha}m_{\alpha}
j_{\beta}m_{\beta}
\middle|JM
\right\rangle
c_{\alpha,m_{\alpha}}^{\dagger}
c_{\beta,m_{\beta}}^{\dagger}
\label{eq:pair_creation_operator}
\end{equation}

In this notation, $\alpha = \{n,l,j,t_z\}$ denotes the quantum numbers
characterizing the single-particle states, while $\epsilon_{\alpha}$
represents the single-particle energy associated with orbital $\alpha$.
The particle-number operator $\hat{n}_{\alpha}$ incorporates the fermion
annihilation and creation operators of the single-particle state,
$c_{\alpha m_{\alpha}}$ and $c^{\dagger}_{\alpha m_{\alpha}}$,
respectively. $A^{\dagger}_{\alpha,\beta}$ is the pair-creation operator,
and $1/\sqrt{1+\delta_{\alpha\beta}}$ represents the normalization factor.
$A_{\gamma,\delta}$ is the Hermitian conjugate of
$A^{\dagger}_{\alpha,\beta}$. The expression
$\langle j_{\alpha}j_{\beta}|V|j_{\gamma}j_{\delta}\rangle_J$
corresponds to the two-body matrix element (TBME) \cite{patel2023systematic}.

In our study, the four $pf$-shell orbits, $1f_{7/2}$, $2p_{3/2}$,
$1f_{5/2}$, and $2p_{1/2}$ orbitals for neutrons as well as protons
are considered as model space with $^{40}$Ca ($N = Z = 20$) serving as the inert doubly magic core. This model space has been extensively tested and validated for nuclear structure studies across the pf model space \cite{poves1981theoretical, caurier2005shell}. We investigate the nuclear structure of even-even Ca isotopes spanning $N = 22$--36 through large scale shell model calculations employing four established $pf$-shell effective interactions: KB3G, GXPF1A, GXPF1B, and FPD6, each constructed through a fundamentally different theoretical approach. The KB3G
interaction \cite{poves2001shell} represents a mass-dependent refinement of the KB3 interaction, which itself is a
monopole-corrected version of the original Kuo-Brown $G$-matrix \cite{kuo1968reaction}. The monopole modifications in
KB3G were applied selectively to the $T = 0$ and $T = 1$ centroids of the  $fp_{1/2}$, $ff_{5/2}$, and $pp$ channels to reproduce simultaneously the empirical quasiparticle gaps at $^{48}$Ca and $^{56}$Ni, with a $(42/A)^{1/3}$ mass dependence
incorporated in the TBMEs. The interaction reduces to KB3 for the $A \leq 52$ nuclei previously studied with the latter and constitutes the standard reference for spectroscopy in the lower pf-shell region \cite{poves2001shell}. The GXPF1A interaction \cite{honma2005shell} is a modification of GXPF1, which was derived from the Bonn-C potential $G$-matrix through least-squares optimization of 70 well-determined linear combinations of TBMEs and single particle energies against approximately 700 experimental energy levels across 87 $pf$-shell nuclei. GXPF1A
was obtained by modifying five TBMEs of GXPF1: three $T = 1$ pairing matrix elements in the $f_{7/2}$ and $p_{1/2}$ sectors were weakened to correct a systematic overestimate of pairing strength arising from the few-dimensional-bases approximation used in the GXPF1 derivation, and two $J = 2, 3$ matrix elements coupling the $f_{5/2}$ and $p_{1/2}$ orbits were adjusted to strengthen the quadrupole-quadrupole interaction in that channel. Each modification lies within approximately  $0.3$--$0.5$ MeV of the original GXPF1 values and collectively improves the description of  $2_1^+$ excitation energies across the calcium, titanium, and chromium isotopic chains.

The GXPF1B interaction \cite{honma2008shell} was subsequently developed from GXPF1A following new spectroscopic data on $^{51,52,53}$Ca reported by Perrot et al. \cite{perrot2006beta}, which imposed additional constraints on TBMEs in the $2p_{3/2}$--$2p_{1/2}$ sector not well-determined in the original GXPF1 fitting procedure. Three modifications distinguish GXPF1B from GXPF1A: the single-particle energy of the  $2p_{1/2}$ orbit is shifted upward by $0.3$ MeV, the $T = 1$ quadrupole-quadrupole interaction strength in the $p_{3/2}$--$p_{1/2}$ channel is increased, and the $p_{1/2}$ pairing matrix elements are readjusted. These changes narrow the $\nu f_{5/2}$--$\nu p_{1/2}$ effective single-particle energy gap toward $N = 34$ relative to GXPF1A, with direct implications for the predicted $N = 34$ subshell closure in $^{54}$Ca.
The parallel use of GXPF1A and GXPF1B in the present calculations permits a direct and quantitative assessment of the role of this monopole correction on both the $N = 32$ and $N = 34$ structural signatures.

The FPD6 interaction \cite{richter1991new} was constructed through a semi-empirical potential approach based on the modified surface one-boson exchange potential (MSOBEP) form, comprising central, spin-orbit, and tensor components with density-dependent and density-independent range terms. Its parameters were determined by iterative least-squares fitting to 61 binding and excitation energies in the mass range $A = 41$ - 49, yielding a final r.m.s. deviation of 176 keV with ten free parameters; the TBMEs incorporate a $(42/A)^{0.35}$ mass dependence. Since FPD6 shares no derivation lineage with the Kuo-Brown $G$-matrix family underlying KB3G and GXPF1(A/B), its inclusion introduces a genuinely independent theoretical constraint on then $N = 32$ subshell closure. The single-particle energies (SPEs) in MeV for all four interactions are compiled in Table~\ref{Table1:SPEs}. 

All calculations are performed with the KSHELL code \cite{shimizu2013nuclear, shimizu2019thick}, which implements M-scheme configuration interaction diagonalization. The many-body wave functions are expressed as linear combinations of Slater determinants defined over the active single-particle orbits, permitting an untruncated treatment of the valence-space correlations. For the heavier calcium isotopes in the present range, the resulting Hamiltonian matrix dimensions reach several tens of millions, and the required lowest eigenvalues and eigenvectors are extracted using the thick-restart block Lanczos algorithm, which combines iterative Krylov-subspace convergence with periodic basis compression to maintain computational tractability at large matrix dimensions. 

The calculations were performed on a high-performance computing
workstation equipped with an Intel Core i7 12th Gen processor with a
maximum clock speed of 4.90~GHz and 32~GB of RAM. The system has 12
physical cores, 20 threads, and 512~GB + 1~TB of storage. The calculations took from 8 hours to 48 hours for the isotopic range of $N = 22$--36.

The excitation energies of the initial and subsequent excited states,
reduced electric quadrupole transition probabilities,
$B(E2: 0_1^+ \rightarrow 2_1^+)$, and quadrupole deformation parameters
($\beta_2$) are computed using the standard effective charges
$e_{\pi} = 1.5e$ and $e_{\nu} = 0.5e$ \cite{bohr1969volume}. These effective charges account phenomenologically for core-polarization contributions from configurations outside the $pf$-shell valence space. The quadrupole deformation parameter $\beta_2$ is calculated using Eq.~\ref{eq:beta2} \cite{ali2024theoretical, raman2001transition, pritychenko2012update,pritychenko2016tables}, given by
\begin{equation}
\beta_2 =
\frac{4\pi}{3ZR_0^2}
\left[
\frac{B(E2)}{e^2}
\right]^{1/2}
\label{eq:beta2}
\end{equation}

where
$R_0 = 1.2A^{1/3}$~fm
represents the nuclear radius, and $Z$ is the atomic number.

\section{Results and discussion }\label{sec:results}
\subsection{Macroscopic Signatures of the $N = 32$ Sub-shell Closure} \label{subsec:macroscopic}
The systematic evolution of six complementary ground-state observables across the even-mass Ca isotopes from $N = 22$ to $N = 36$ is examined in the present section. The experimental data adopted from the NNDC database \cite{NNDC_NuDat3}, the charge-radius compilation of Angeli and Marinova \cite{angeli2013table}, and the ADNDT (2021) tabulation \cite{li2021compilation} are compared with predictions from seven theoretical frameworks; RCHB (2018), FRDM (2012), dHRB (2024), and the four RHB parametrizations PC-L3R, PC-X, DD-MEX, and DD-PCX spanning relativistic mean-field, deformed Hartree-Bogoliubov, and macroscopic-microscopic theoretical approaches. The observables are presented in order of increasing sensitivity to subshell structure, and the discussion focuses on the identification of model-independent signatures of the $N = 32$ sub-shell closure across this multi-framework ensemble.

\subsubsection{Two neutron separation energies} 
\label{subsubsec:s2n}
The two-neutron separation energies $S_{2n} (N, Z)$ computed from Eq.~\ref{eq:S2n} are displayed in Fig. \ref{eq:S2n} as a function of neutron number for all seven theoretical frameworks alongside the NNDC experimental data. All frameworks reproduce the overall monotonically decreasing trend across the isotopic chain, reflecting the systematic reduction in two-neutron binding as the neutron Fermi level approaches the top of the $pf$-shell. A pronounced change in the rate of decrease is observed at $N = 28$ in both the experimental data and all theoretical calculations, consistent with the well-established major shell closure at this neutron number and the opening of the $\nu p_{1/2}$ subshell above $^{48}$Ca.  Remarkably, the NNDC experimental data exhibit a structurally analogous, albeit weaker, feature at higher neutron number: a step-like reduction of $S_{2n}$ values at $N = 32$ mirroring the qualitative behavior observed at the $N = 28$ closure and consistent with the emergence of a sub-shell gap at $N = 32$. This secondary step feature is absent from the theoretical frameworks, which display a smooth and monotonic descent through this region without capturing the local structural retardation evident in the experimental data. The insensitivity of the theoretical $S_{2n}$ predictions to this feature reflects the limited resolution of absolute two-neutron separation energies as probes of sub-shell structure and motivates the systematic examination of higher-order differential quantities presented in Figs. \ref{fig:dS2n}, \ref{fig:delta1nBE}, and \ref{fig:deltaE}, which are specifically constructed to amplify such localized discontinuities in the binding energy surface.

\subsubsection{Differential two-neutron separation energies:} \label{subsubsec:ds2n}
The differential separation energies $dS_{2n}$ defined in Eq.~\ref{eq:dS2n}, shown in Fig. \ref{fig:dS2n}, amplify the sensitivity to shell structure by extracting the local curvature of the $S_{2n}$ surface. A dominant minimum is observed at $N = 28$ in both the experimental data and all seven theoretical frameworks, unambiguously confirming the major shell closure at this neutron number. Of particular significance is the behavior at $N = 32$, the NNDC experimental data exhibits a clear secondary minimum at this neutron number, distinct from the smooth background trend between $N = 30$ and $N = 36$. This feature, while less pronounced than the $N = 28$ signal by approximately a factor of three in depth, is qualitatively reproduced by RCHB, dHRB, PC-L3R, PC-X, DD-MEX, and DD-PCX, providing initial macroscopic evidence for enhanced neutron binding at $N = 32$. The FRDM (2012) framework does not reproduce the $N = 32$ feature in $dS_{2n}$, exhibiting a monotonic recovery from the $N = 28$ minimum that is inconsistent with the experimental trend. This anomalous behavior of FRDM, which persists across several observables discussed below, likely reflects the limitations of the macroscopic-microscopic approach in capturing the fine monopole structure of the single-particle level ordering in this mass region.

\subsubsection{Three-point binding energy filter} 
\label{subsubsec:3-point-be}
The three-point binding energy filter $\Delta_{1n}^{(3)} B.E. (N, Z)$ defined in Eq.~\ref{eq:three_point_mass_filter}, presented in Fig. \ref{fig:delta1nBE}, provides a considerably sharper probe of shell structure by suppressing the smooth mass-dependent background of the $S_{2n}$ surface. A pronounced global minimum at $N = 28$ is reproduced by all frameworks, confirming the dominant shell effect at this neutron number. Proceeding to higher neutron numbers, a sharp local maximum at $N = 30$ is followed by a well-defined secondary minimum at $N = 32$ in the adopted NNDC experimental results, with a depth that approaches that of the$N=28$ feature. This secondary minimum is reproduced qualitatively by RCHB, dHRB, DD-MEX, DD-PCX, PC-L3R, and PC-X, which all exhibit a local minimum at $N = 32$ of varying depth, while FRDM again behaves anomalously, displaying a broad and featureless variation in this region that fails to capture the $N = 32$ structural discontinuity. The convergence of six independent relativistic and deformed Hartree-Bogoliubov frameworks on a consistent $N = 32$ signature in this observable constitutes strong macroscopic evidence that the enhanced stability at this neutron number is a robust physical feature rather than a model-dependent prediction.

\subsubsection{Neutron shell gap energies} 
\label{subsubsec:delta-e}
The most compelling binding energy evidence for the $N = 32$ sub-shell closure is provided by the shell gap energy $\Delta E$ defined in Eq. \ref{eq:neutron_shell_gap}, shown in Fig. \ref{fig:deltaE}. This second-order finite difference of the binding energy directly extracts the magnitude of the shell gap and is the most sensitive of the bulk observables employed in the present analysis. The adopted experimental NNDC results displays a dominant minimum at $N = 28$ of $-3.54$ MeV, confirming the major shell closure, followed by a recovery and a well-resolved secondary minimum at $N = 32$ reaching approximately $-2.18$ MeV. This $N = 32$ feature in the experimental data is the most pronounced sub-shell closure signature observed across all six macroscopic observables studied here. Among the theoretical frameworks, RCHB, dHRB, PC-L3R, PC-X, DD-MEX, and DD-PCX all exhibit a local minimum at $N = 32$, qualitatively consistent with the experimental observation, though all underestimate the depth of the experimental signal to varying degrees. DD-MEX produces the deepest theoretical minimum at $N = 32$ among the frameworks, while FRDM (2012) again fails to reproduce the feature, displaying a monotonic approach to zero that is inconsistent with both the experimental data and the other theoretical predictions. The systematic underestimation of the $N = 32$ gap depth by the theoretical frameworks relative to experiment may reflect the incomplete treatment of beyond-mean-field correlations and the limited sensitivity of these approaches to the monopole tensor interaction responsible for the sub-shell closure.

\subsubsection{Nuclear rms charge radii} 
\label{subsubsec:rms-radii}
The root-mean-square charge radii $r_{{ch}}$  Eq. \ref{eq:charge_radius}, are shown in Fig. \ref{fig:r_ch}, where the nuclear charge radius $(R_C)$ incorporating shell  $(\Delta E)$ and deformation $(\beta_2, \beta_4)$ corrections is compared against the experimental values from Angeli and Marinova, available for $N = 22-28$ and ADNDT (2021), available for $N = 23-32$. All theoretical frameworks reproduce the overall increasing trend of $r_{ch}$ with neutron number across the available experimental range, reflecting the systematic expansion of the nuclear charge distribution as neutrons progressively fill the $pf$-shell. The theoretical ensemble maintains good quantitative agreement with the Angeli and Marinova data throughout the isotopic chain from $N = 22$ to $N = 32$, validating the charge radius parametrization employed. The pronounced minimum in $r_{ch}$ at $N = 28$, reproduced consistently across the theoretical ensemble and both experimental datasets, confirming the magicity at $N = 28$. A secondary minimum of smaller magnitude at $N = 32$ is evident in the NNDC data and most theoretical frameworks, consistent with a local reduction in collectivity associated with the $N = 32$ sub-shell closure. A shallower minimum at $N = 34$ is discernible in several frameworks, suggesting a residual structural effect at this neutron number, though its magnitude is considerably less pronounced than that observed at $N = 32$.

\subsubsection{Three-point charge radius filter: } \label{subsubsec:3-point-radii}
The three-point charge radius filter $\Delta_{1n}^{(3)}r_{ch}(N,Z)$ defined in Eq.~\ref{eq:three_point_charge_radius}, displayed in Fig. \ref{fig:delta1nRch}, reveals the shell structure information embedded in the charge radius evolution with considerably greater resolution than the rms charge radius values. The theoretical frameworks, together with the calculated results from NNDC and available experimental results i.e, Angeli and Marinova and ADNDT (2021), exhibits a deep minimum at $N = 28$, confirming the well-established shell closure. A secondary minimum at $N = 32$ is clearly present in the NNDC results, providing direct evidence on the $N = 32$ subshell closure. Most of the theoretical frameworks reproduces this secondary minimum qualitatively, while FRDM again displays anomalous behavior throughout the neutron-rich region. It is noted that the $\Delta_{1n}^{(3)}r_{ch}(N,Z)$ observable is particularly significant in the context of the present study, as it is directly related to the charge radius measurements of Koszorús et al. \cite{koszorus2021charge} in potassium isotopes that originally motivated this comprehensive investigation.

The systematic analysis of macroscopic observables consistently indicates enhanced structural stability at $N = 32$ in the calcium isotopic chain, with six out of seven theoretical frameworks reproducing the closure signatures across multiple binding energy and charge radius quantities in qualitative agreement with the experimental data, while FRDM remains a notable exception; the single-particle origin of these macroscopic features is examined through the RMF analysis in Sec.~\ref{subsec:SPE-RMF}.

\subsection{Single-Particle energy levels from Relativistic Mean-Field Analysis } \label{subsec:SPE-RMF}
\begin{figure}
    \centering
    \includegraphics[width=0.75\textwidth]{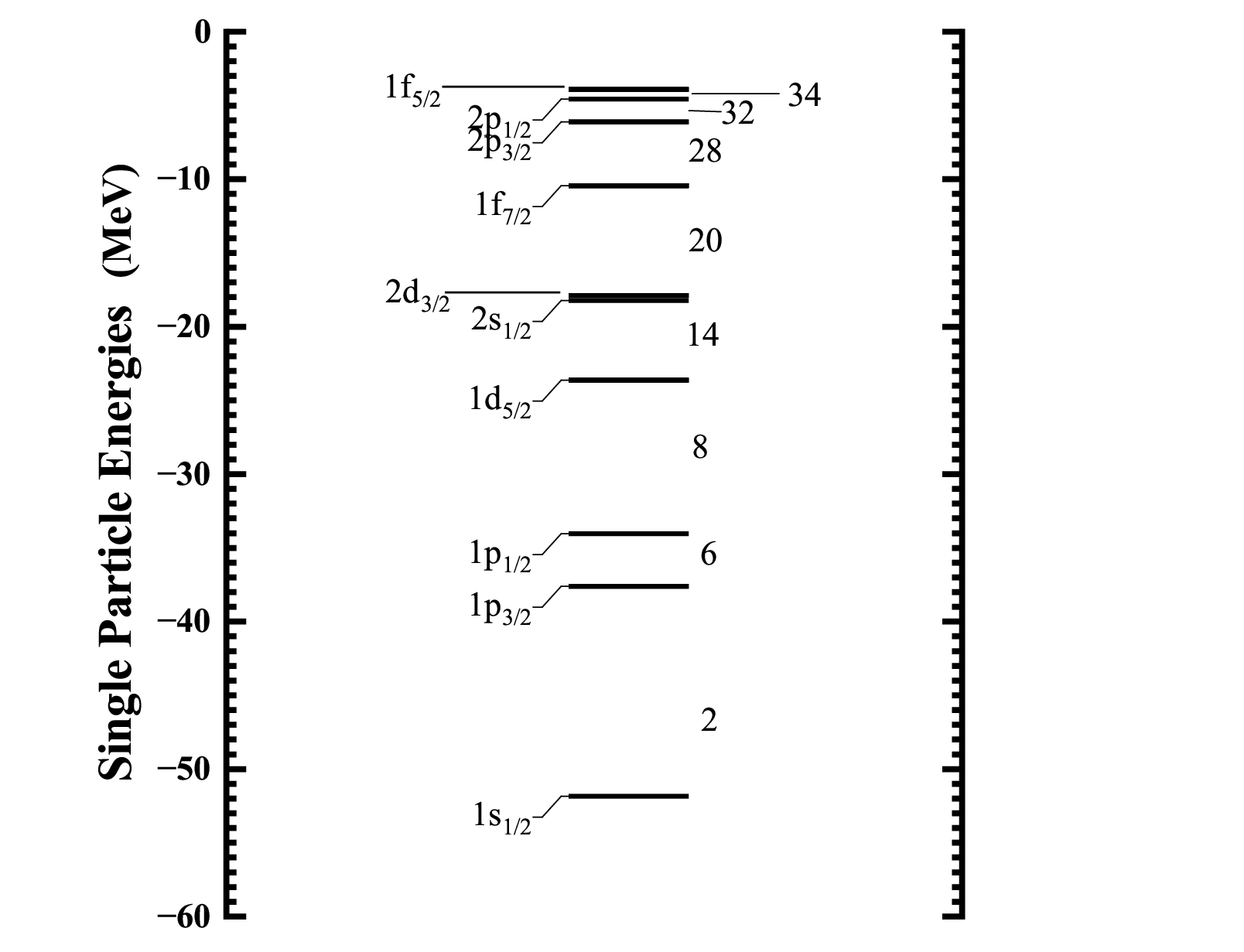}
    \caption{Single particle energy levels of $^{56}$Ca isotope, estimated for RMF model with NL3* parameter set.}
    \label{fig:SPEsCa}
\end{figure}

The neutron single-particle energy spectrum of $^{56}$Ca, computed
within the RMF framework using the NL3* force parameter set, is presented
in Fig. \ref{fig:SPEsCa}. The calculated level scheme reproduces the established deformed shell/sub-shell closures $2$, $6$, $8$, $14$, $20$, and $28$ as prominent energy gaps in the
single-particle spectrum, consistent with the well-known predictive
capability of the relativistic mean-field approach for closed-shell
nuclei \cite{gambhir1990relativistic, lalazissis1997new, swain2018nuclear}. Within the $pf$-shell region, the $1f_{7/2}$ orbital
lies well separated from the remaining $pf$-shell orbits, accounting for
the $N = 28$ shell closure above the $^{40}$Ca core. Of particular
relevance to the present investigation is the level ordering above
$N = 28$. The $2p_{3/2}$ orbital is clearly separated from the nearly
degenerate $2p_{1/2}$ and $1f_{5/2}$ cluster, with a visible energy gap
between $2p_{3/2}$ and $2p_{1/2}$ that corresponds to the $N = 32$
subshell closure identified in the macroscopic analysis of
Sec.~\ref{subsec:macroscopic}. A smaller but discernible separation
between $2p_{1/2}$ and $1f_{5/2}$ is also apparent, consistent with the
weaker $N = 34$ structural features observed in the charge-radius
systematics. The single-particle level ordering predicted by NL3* thus
provides a direct mean-field interpretation of the macroscopic shell
closure signatures, indicating that the $N = 32$ deformed sub-shell gap arises from the energy spacing between the $2p_{3/2}$ and $2p_{1/2}$ orbitals within the self-consistent relativistic potential.

\subsection{Microscopic Shell Model Calculations } 
\label{subsec:micro-shell-model} 
The spectroscopic results derived from shell-model (SM) calculations
for even-even Ca isotopes with $N = 22$--36 are discussed in this section.
The excitation energies of the yrast states are compared with the
corresponding experimental data adopted from NNDC and previous theoretical studies in
Sec.~\ref{subsubsec:e2}. The reduced $E2$ transition probabilities and
deformation parameters of the even-even Ca isotopes are investigated in
Sec.~\ref{subsubsec:Be2}.

\subsubsection{Energy Spectra } 
\label{subsubsec:e2}
\begin{figure}
    \centering
    \includegraphics[width=0.75\textwidth]{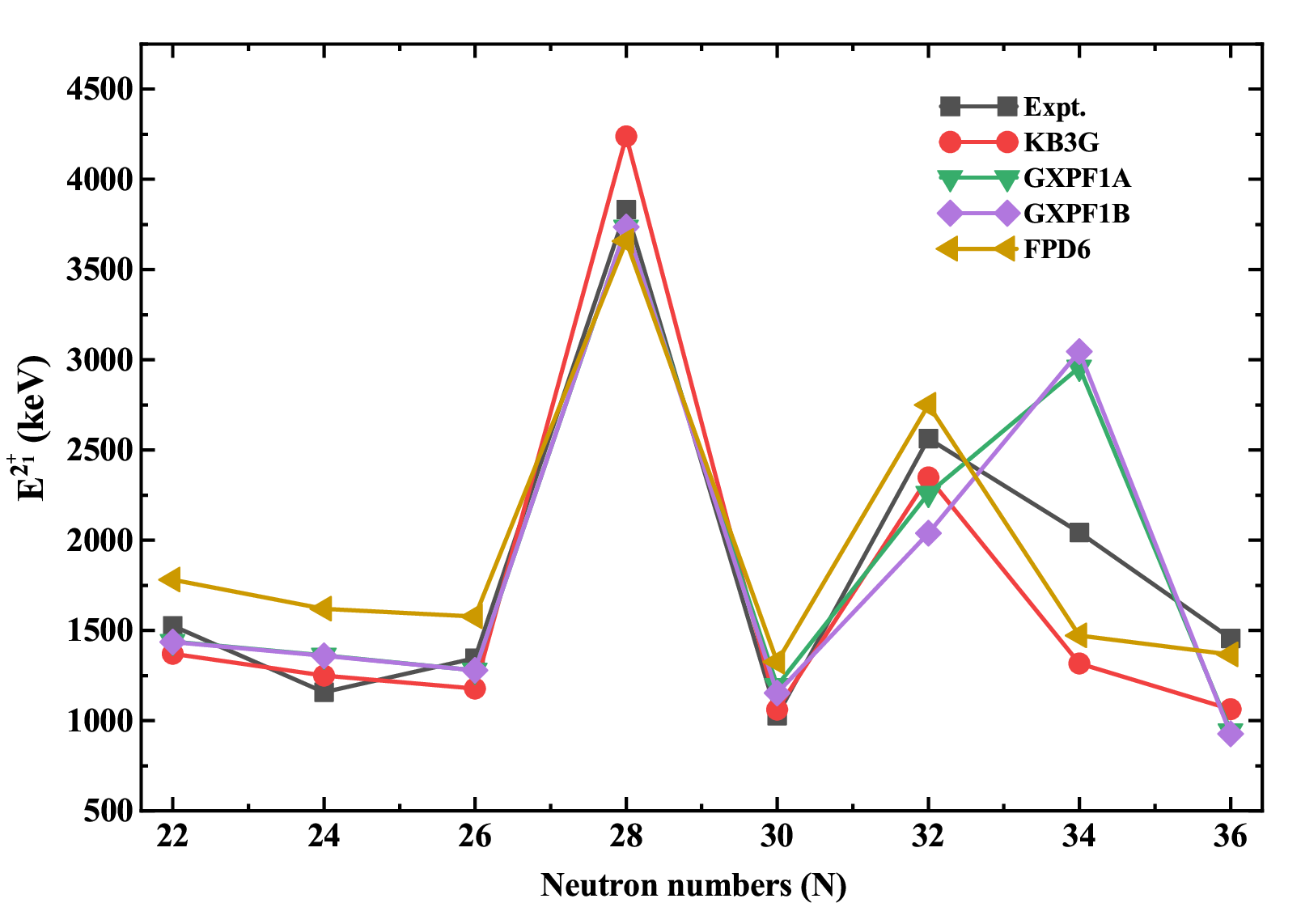}
    \caption{Comparison of $E(2_1^+)$ (in keV) obtained from SM calculations using the KB3G, GXPF1A, GXPF1B, and FPD6 interactions with the experimental datas \cite{NNDC_NuDat3, ali2024theoretical, raman2001transition, pritychenko2012update, pritychenko2016tables, CHEN20161, CHEN20231, WU20001, CHEN20221, CHEN20191, DONG2015185, DONG20141, chen2023level} for even--even $^{42-56}$Ca isotopes, except for $^{54}$Ca ($N=34$), where the $2_1^+$ energy taken from Steppenbeck et al. \cite{steppenbeck2013evidence}.}
    \label{fig:E2}
\end{figure}

\begin{figure}
    \centering
    \includegraphics[width=0.75\textwidth]{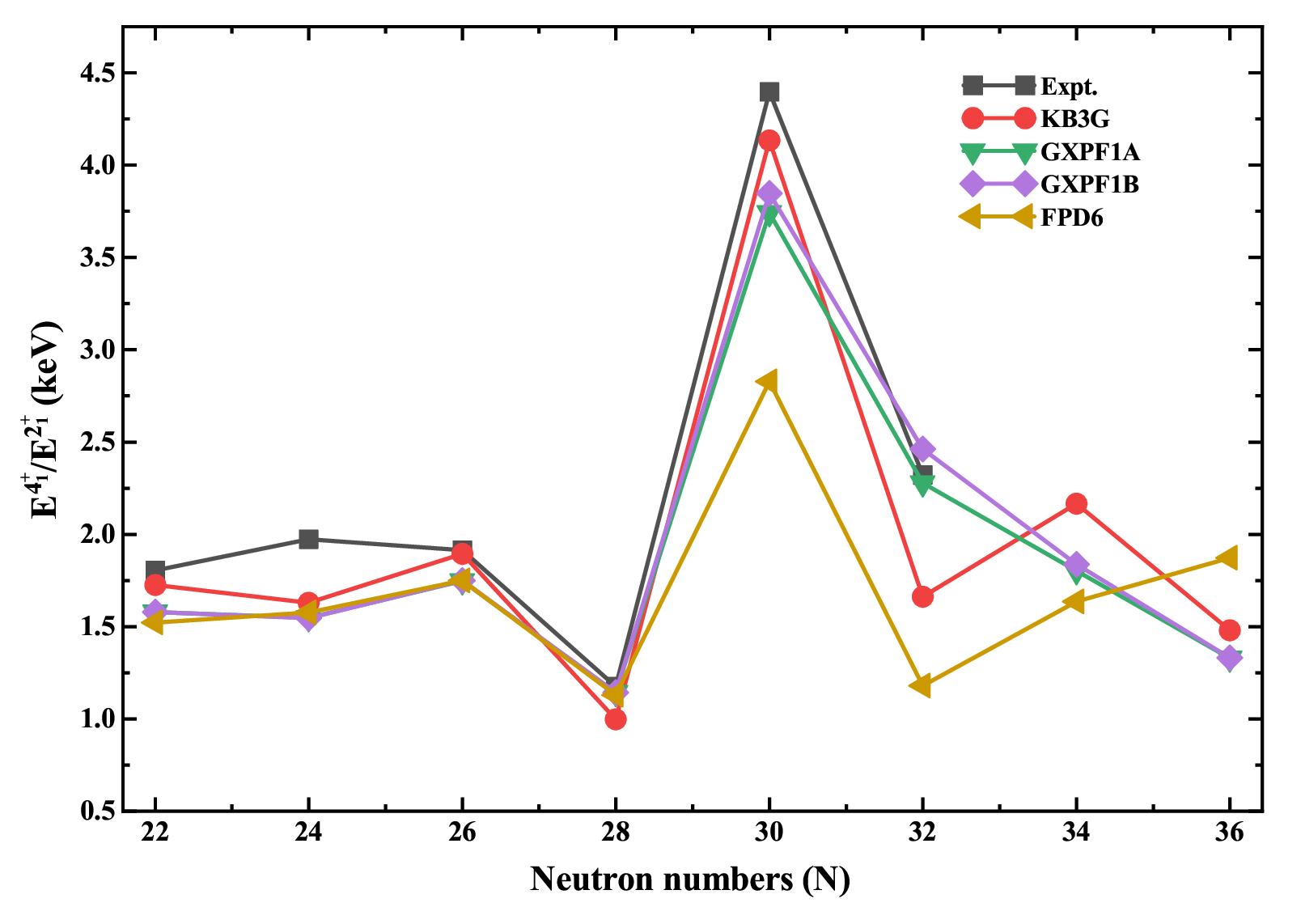}
    \caption{Comparison of the $R_{4/2}=E(4_1^+)/E(2_1^+)$ ratio obtained from SM calculations using the KB3G, GXPF1A, GXPF1B, and FPD6 interactions with the experimental datas \cite{NNDC_NuDat3, steppenbeck2013evidence, ali2024theoretical, raman2001transition, pritychenko2012update, pritychenko2016tables, CHEN20161, CHEN20231, WU20001, CHEN20221, CHEN20191, DONG2015185, DONG20141, chen2023level}  for even--even $^{42-56}$Ca isotopes.}
    \label{fig:R4by2}
\end{figure}

The energy ratio $R_{4/2}$, defined as
$R_{4/2} = \frac{E(4_1^+)}{E(2_1^+)}$
constitutes an important observable for analyzing collective nuclear properties. This parameter provides insight into the nature of the lowest-lying $2_1^+$ and $4_1^+$ excitations. While the yrast $2_1^+$ state generally represents a member of the ground-state band in even-even nuclei, the $4_1^+$ state may not necessarily belong to the same rotational sequence \cite{sahoo2024nuclear}. In Figs. \ref{fig:E2} and \ref{fig:R4by2}, the calculated lowest excited-state energies, $E(2_1^+)$, and the corresponding $R_{4/2}$ ratios for the even-even Ca isotopes from $N=22$ to $N=36$ are compared with the available experimental data \cite{NNDC_NuDat3, steppenbeck2013evidence, ali2024theoretical, raman2001transition, pritychenko2012update, pritychenko2016tables, CHEN20161, CHEN20231, WU20001, CHEN20221, CHEN20191, DONG2015185, DONG20141, chen2023level}. The present calculations generally show good agreement with the experimental trends across the isotopic chain.

All four interactions reasonably reproduce the overall trend of $E(2_1^+)$, with GXPF1A and GXPF1B generally yielding slightly higher excitation energies than KB3G and FPD6. The pronounced peak at $N=28$, obtained consistently across all interactions and in agreement with experiment, confirms the robust shell closure at $^{48}$Ca. This behavior is associated with the large shell gap accompanying the completion of the neutron $1f_{7/2}$ orbital, which suppresses low-energy particle-hole excitations across the gap and favors a predominantly spherical structure. Beyond $N=28$, the sharp reduction in $E(2_1^+)$ at $N=30$, as neutrons begin occupying the $2p_{3/2}$ orbital, marks the onset of enhanced quadrupole collectivity within the $pf$ shell. This behavior is accompanied by an increase in the $R_{4/2}$ ratio, approaching $\approx 4.0$ at $N=30$, indicative of enhanced quadrupole correlations and a more rotational-like level spacing. For comparison, the ideal axially symmetric rigid-rotor limit is $R_{4/2}=3.33$ \cite{abdullah2014b}.

The behavior at $N=32$ distinguishes the four interactions. KB3G and FPD6 show a modest rise in $E(2_1^+)$ relative to $N=30$, accompanied by a reduction in $R_{4/2}$, consistent with a partial suppression of collectivity and the emergence of a subshell gap, as reported in previous experimental and theoretical studies \cite{wienholtz2013masses, garcia2015ground, bhoy2020shell, akkoyun2020nuclear, enciu2022extended, li2024spectroscopy, liu2026shell, anupam2023shell}. This gap is associated with the completion of the neutron $2p_{3/2}$ orbital at $N=32$, producing an energy separation to the $2p_{1/2}$ orbital. The resulting reduction in the number of available low-energy configurations stabilizes the nuclear structure at $N=32$, although the effect is weaker than that observed at $N=28$. Experimental $R_{4/2}$ values are not available beyond $N=30$ for the isotopes considered here, owing to the lack of experimentally established $4_1^+$ excitation energies; therefore, the calculated ratios provide predictions that can be tested by future measurements.

In contrast, GXPF1A and GXPF1B show no clear enhancement at $N=32$, instead exhibiting a more gradual evolution toward $N=34$. This difference reflects the distinct monopole evolution of the GXPF1-family interactions, which favors an enhanced separation between the neutron $2p_{1/2}$ and $1f_{5/2}$ orbitals toward $N=34$. Consequently, the predicted shell evolution differs from that obtained with KB3G and FPD6, leading to a weaker $N=32$ signature and a more pronounced evolution toward $N=34$. This highlights the sensitivity of the predicted subshell structure to the monopole component of the effective interaction. At $N=36$, both GXPF1A and GXPF1B predict a marked reduction in the excitation energy, suggesting a resurgence of collectivity toward the neutron-rich end of the investigated isotopic chain. In the absence of corresponding experimental data, this prediction remains to be tested experimentally.

\subsubsection{Electromagnetic Properties} 
\label{subsubsec:Be2}
We extend the analysis to the electromagnetic properties of the even-even Ca isotopes from $N=22$ to $N=36$ using all four effective interactions within the KSHELL framework. Table \ref{Table2:BE2} presents the calculated $B(E2; 0_1^+\rightarrow2_1^+)$ transition strengths and the derived deformation parameters ($\beta_2$), together with the available experimental data \cite{NNDC_NuDat3, raman2001transition, pritychenko2012update, pritychenko2016tables} . These quantities test the wave functions of the effective interactions, which the excitation energies alone do not constrain, and they trace the evolution of quadrupole collectivity from the $N=28$ shell closure into the neutron-rich region, where experimental data remain limited.

\begin{figure}
    \centering
    \includegraphics[width=0.75\textwidth]{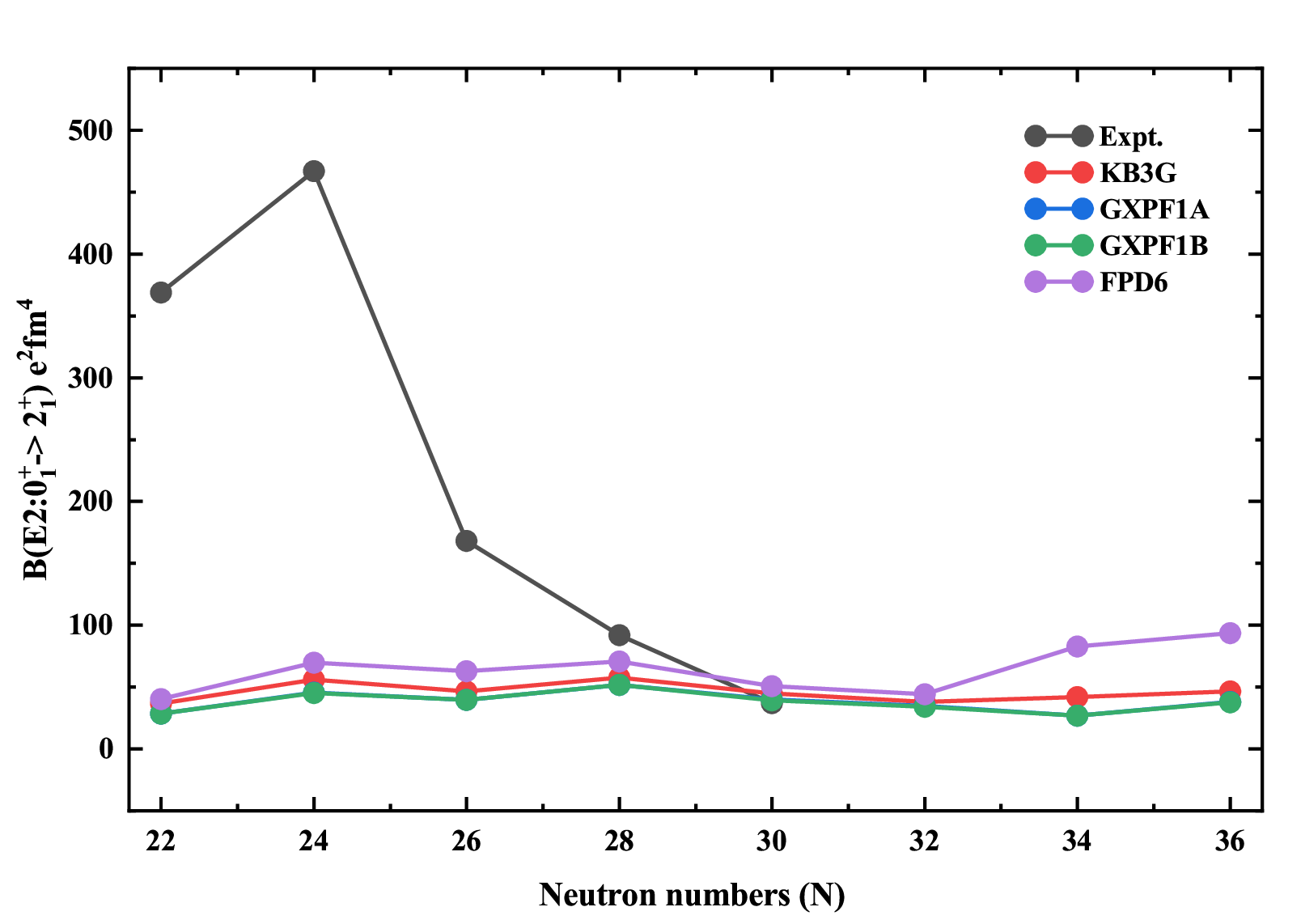}
    \caption{Comparison of $B(E2)$ transition strengths (in $e^2$fm$^4$) obtained from SM calculations using the KB3G, GXPF1A, GXPF1B, and FPD6 interactions with the experimental data \cite{NNDC_NuDat3, raman2001transition, pritychenko2012update, pritychenko2016tables} for even-even $^{42-56}$Ca isotopes.}
    \label{fig:B(E2)}
\end{figure}

The calculations use the standard effective charges $e_\pi=1.5e$ and $e_\nu=0.5e$. Because $Z=20$ leaves no valence protons above the $^{40}$Ca core, only the neutron effective charge $e_\nu$ contributes to the calculated transition matrix element, and $B(E2)$ scales as $e_\nu^2$. The calculated values lie between about 27 and 94 $e^2$fm$^4$ and vary weakly with neutron number, with FPD6 giving the largest values and GXPF1A and GXPF1B giving nearly identical results. In contrast, the experimental values decrease from $B(E2)=467$ $e^2$fm$^4$ at $N=24$ ($^{44}$Ca) to $37.3$ $e^2$fm$^4$ at $N=30$ ($^{50}$Ca), a reduction by a factor of about 12.5 (Fig. \ref{fig:B(E2)}), whereas the calculated values decrease by only 12 to 27\% over the same interval.

The largest discrepancies occur at $N=22$ to $26$, where experiment exceeds the calculations by factors of 9 to 13 at $N=22$, 7 to 10 at $N=24$, and about 3 to 4 at $N=26$. This underestimation is attributed to the restriction to the $pf$ shell. In this region, cross-shell particle-hole excitations across the $sd$-$pf$ gap and contributions from higher-lying orbitals beyond the $pf$ model space can enhance the quadrupole collectivity, and these effects are not included in the present model space. Since a uniform rescaling of $e_\nu$ changes only the overall magnitude, it cannot reproduce the observed neutron-number dependence. The disparity decreases with increasing neutron number. At $N=28$ the calculations underestimate the experimental value of 92 $e^2$fm$^4$ by 23 to 44\%, whereas at $N=30$ they overestimate the experimental value of 37.3 $e^2$fm$^4$ by 6 to 36\%. For KB3G, GXPF1A and GXPF1B the deviation is smaller at $N=30$ than at $N=28$, while FPD6 deviates less at $N=28$ (23\% against 36\%). Because the sign of the deviation changes between the two isotopes, no single value of $e_\nu$ would reproduce both. The agreement near $N=30$ is therefore interaction dependent, and two isotopes do not establish a systematic trend.

\begin{figure}
    \centering
    \includegraphics[width=0.75\textwidth]{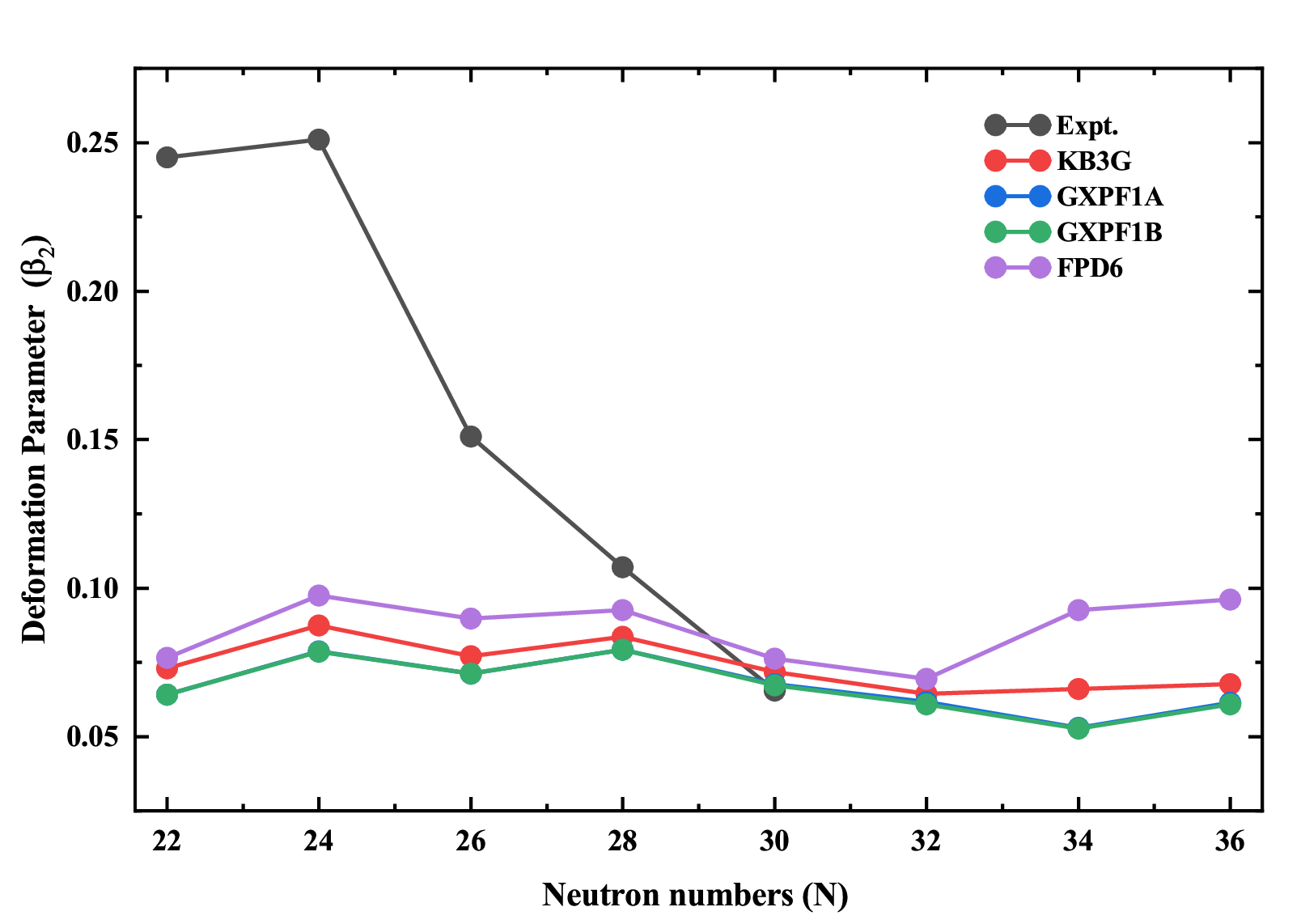}
    \caption{Comparison of deformation parameter ($\beta_2$) obtained from SM calculations using the KB3G, GXPF1A, GXPF1B, and FPD6 interactions with the experimental data \cite{NNDC_NuDat3, raman2001transition, pritychenko2012update, pritychenko2016tables} for even-even $^{42-56}$Ca isotopes.}
    \label{fig:beta2}
\end{figure}

The deformation parameter $\beta_2$ derived from $B(E2)$ reflects the same
systematics (Fig. \ref{fig:beta2}). The calculated values remain between
0.05 and 0.10 along the isotopic chain, whereas the experimental values
decrease from $\beta_2\simeq0.25$ at $N=22$ and 24 to
$\beta_2\simeq0.065$ at $N=30$. Over the same interval, the calculated
values decrease by only 14 to 22\%. This discrepancy can be attributed,
at least in part, to the restriction to the $pf$ shell, since cross-shell
excitations are known to enhance the quadrupole collectivity of the
lighter Ca isotopes \cite{garcia2015ground, li2024spectroscopy, liu2026shell,  coraggio2009spectroscopic, caurier2001shell, gade2006cross, holt2014three,
ash2021cross}.


The agreement between calculation and experiment is closest at $N=30$,
but it is not systematic. At $N=28$, the calculated $\beta_2$ values lie
14 to 26\% below the experimental value, whereas at $N=30$ they lie 3 to
17\% above it. Thus, agreement at these two isotopes alone does not
establish the reliability of the four interactions over the $N=30$--36
region, where the experimental compilations used here provide no
$B(E2)$ values.

For $N>30$, the predictions from the four interactions diverge. All four
give a $B(E2)$ that is 13 to 15\% below the $N=30$ value at $N=32$.
GXPF1A and GXPF1B then reach a minimum at $N=34$
($\beta_2\simeq0.053$), coinciding with the maximum of
$E(2_1^+)$ in Fig. \ref{fig:E2}, and return to within 5\%
of their $N=30$ values at $N=36$. In contrast, FPD6 predicts an increase
of $B(E2)$ by 63\% at $N=34$ and 85\% at $N=36$, in a region outside the
mass range $A=41$--49 to which it was fitted \cite{richter1991new}.
The predicted quadrupole strength for $N>30$ is therefore strongly
interaction dependent, and measurements of $B(E2)$ in this region would
provide a direct experimental test of the different predictions.

\section{Conclusions}
\label{sec:conclusions}

In this work, we have investigated the structural evolution of even-even
Ca isotopes from $N=22$ to 36 by combining complementary macroscopic,
mean-field, and large-scale shell-model approaches. The analysis includes
ground-state observables, neutron shell-gap indicators, charge-radius
systematics, relativistic mean-field (RMF) single-particle energies, and
spectroscopic properties calculated in the full $pf$-shell model space
with the KSHELL code. The shell-model calculations employ the KB3G,
GXPF1A, GXPF1B, and FPD6 effective interactions, allowing the sensitivity
of the predicted shell evolution to the underlying interaction to be
examined explicitly.

The macroscopic observables provide mutually consistent evidence for a
change in the neutron structure at $N=32$. In particular, the neutron
shell-gap indicator $\Delta E$ and the three-point binding-energy filter $\Delta_{1n}^{(3)} B.E. (N, Z)$ exhibit the
most pronounced signatures, while the three-point charge-radius filter $\Delta_{1n}^{(3)}r_{ch}(N,Z)$ shows a corresponding change in the evolution of the nuclear charge distribution. The agreement among these complementary observables is significant because they probe different aspects of the ground-state systematics. Taken together, they provide a consistent indication of
enhanced structural stability at $N=32$. The absence of a comparable
signature in the FRDM results further illustrates the sensitivity of the
predicted shell evolution to the treatment of microscopic single-particle
and monopole effects in this mass region.

The RMF calculations for $^{56}$Ca with the NL3$^*$ parameterization provide a single-particle interpretation of this behavior. A pronounced separation between the neutron $2p_{3/2}$ and $2p_{1/2}$ orbitals develops around $N=32$, consistent with the shell-gap signatures obtained from the
macroscopic analysis. Within the present RMF description, this
single-particle evolution provides a microscopic basis for the enhanced
stability observed at $N=32$ and connects the ground-state signatures to
the underlying neutron orbital structure.

The shell-model results establish a complementary spectroscopic picture.
The conventional $N=28$ shell closure is robustly reproduced by all four
interactions, as reflected by the pronounced maximum of
$E(2_1^+)$ and the corresponding change in $R_{4/2}$. Beyond $N=28$, the
interactions predict different degrees of quadrupole collectivity. In
particular, KB3G and FPD6 reproduce the enhancement of $E(2_1^+)$ at
$N=32$ together with the corresponding reduction of $R_{4/2}$, yielding a
qualitative signature consistent with the available experimental data.
GXPF1A and GXPF1B, in contrast, give a weaker $N=32$ signature and a more
pronounced evolution toward $N=34$. This difference reflects the distinct
monopole evolution incorporated in the interactions and demonstrates the
sensitivity of the detailed shell evolution to the effective interaction.

The electromagnetic observables provide an additional constraint on the
calculated wave functions. The experimental $B(E2)$ values and the derived
$\beta_2$ decrease strongly from $N=22$--24 toward $N=30$, whereas the
restricted $pf$-shell calculations show a substantially weaker variation.
This discrepancy is consistent with the omission of cross-shell
particle-hole excitations from the present $^{40}$Ca-core model space,
which can contribute significantly to quadrupole collectivity in the
lighter Ca isotopes. The agreement between theory and experiment becomes
closer around $N=28$--30, but this agreement is not sufficient to establish
the reliability of the interactions throughout the neutron-rich region,
particularly because experimental electromagnetic data become sparse
beyond $N=30$.

The predictions beyond $N=30$ are consequently more interaction dependent.
All four interactions predict a reduction of $B(E2)$ at $N=32$ relative to
$N=30$, while GXPF1A and GXPF1B continue to a minimum around $N=34$ before
recovering toward $N=36$. FPD6 instead predicts a substantial increase in
$B(E2)$ beyond $N=30$. The corresponding differences in $\beta_2$ and
$E(2_1^+)$ demonstrate that the predicted evolution of quadrupole
collectivity beyond $^{50}$Ca is not uniquely determined within the
present set of interactions. Measurements of electromagnetic transition
strengths in $^{52,54,56}$Ca would therefore provide an important
experimental constraint on the competing predictions.

Overall, the combined analysis supports $N=32$ as a robust subshell
feature in the Ca isotopic chain. Its signature is observed consistently
in several ground-state observables, is supported by the RMF
single-particle evolution of the neutron $2p_{3/2}$ and $2p_{1/2}$ orbitals,
and is reflected in the spectroscopic behavior obtained with selected
shell-model interactions. These findings are
consistent with recent experimental and theoretical studies of shell
evolution in neutron-rich Ca isotopes \cite{garcia2015ground, li2024spectroscopy, koszorus2021charge}. At the same time, the different predictions
for the evolution toward $N=34$ demonstrate that the detailed location
and strength of subshell effects remain sensitive to the monopole
properties of the effective interaction. The present results therefore
provide a unified description of the available evidence for shell
evolution in neutron-rich Ca isotopes while identifying the electromagnetic
and spectroscopic observables beyond $N=30$ that are most important for
discriminating among competing theoretical descriptions.

\section*{Acknowledgements}

The authors would like to express their gratitude to Dr.~P.~Parida and the HPC group at the Central Research Facility, KIIT-DU, for providing the computational facilities used to perform the LSSM calculations. The authors also sincerely thank Dr.~Noritaka Shimizu of the Center for Computational Sciences, University of Tsukuba, for his invaluable guidance and insightful discussions concerning the implementation and interpretation of the shell-model calculations.





\bibliographystyle{elsarticle-num} 
\bibliography{Ca-article}







\clearpage
\onecolumn
\begin{table}[htbp]
\centering
\begin{tabular}{l c c c c}
\hline\hline
\textbf{Interactions} & \multicolumn{4}{c}{\textbf{SPEs (MeV)}} \\
\hline
 & $1f_{7/2}$ & $2p_{3/2}$ & $1f_{5/2}$ & $2p_{1/2}$ \\
\cline{2-5}
KB3G   & $-8.6000$ & $-6.6000$ & $-2.1000$ & $-4.6000$ \\
GXPF1A & $-8.6240$ & $-5.6793$ & $-1.3829$ & $-4.1370$ \\
GXPF1B & $-8.6240$ & $-5.6793$ & $-1.3829$ & $-3.8370$ \\
FPD6   & $-8.3876$ & $-6.4952$ & $-1.8966$ & $-4.4783$ \\
\hline\hline
\end{tabular}
\caption{Single-particle energies (SPEs), in MeV, used for the effective interactions.}
\label{Table1:SPEs}
\end{table}

\begin{table}[htbp]
\centering
\begin{tabular}{c c c c c c | c c c c c}
\hline\hline
 & \multicolumn{5}{c}{\textbf{$B(E2)$ (e$^2$ fm$^4$)}} &
 \multicolumn{5}{c}{\textbf{$\beta_2$}} \\
\textbf{$N$} & \textbf{NNDC} & \textbf{KB3G} & \textbf{GXPF1A} &
\textbf{GXPF1B} & \textbf{FPD6} & \textbf{NNDC} & \textbf{KB3G} &
\textbf{GXPF1A} & \textbf{GXPF1B} & \textbf{FPD6} \\
\hline
22 & 369   & 36.8 & 28.4 & 28.4 & 40.4 & 0.245 & 0.07302 & 0.06415 & 0.064149 & 0.07651 \\
24 & 467   & 56.1 & 45.4 & 45.3 & 69.7 & 0.251 & 0.0874  & 0.07863 & 0.078543 & 0.097427 \\
26 & 168   & 46.3 & 39.5 & 39.5 & 62.8 & 0.151 & 0.07708 & 0.0712  & 0.071202 & 0.089778 \\
28 & 92    & 57.6 & 51.8 & 51.7 & 70.7 & 0.107 & 0.08357 & 0.07925 & 0.07918  & 0.092593 \\
30 & 37.3  & 44.8 & 39.8 & 39.4 & 50.6 & 0.0654 & 0.07173 & 0.0676 & 0.067266 & 0.07623 \\
32 & --    & 38.1 & 34.8 & 34.0 & 44.2 & -- & 0.06444 & 0.06158 & 0.060874 & 0.069407 \\
34 & --    & 42.0 & 27.0 & 26.8 & 82.6 & -- & 0.06597 & 0.0529 & 0.052703 & 0.092525 \\
36 & --    & 46.4 & 38.1 & 37.5 & 93.6 & -- & 0.06768 & 0.06133 & 0.060849 & 0.096134 \\
\hline\hline
\end{tabular}
\caption{Comparison of SM calculated and experimental $B(E2)$ transition
strengths and deformation parameter $\beta_2$ for even--even
$^{42-56}$Ca isotopes.}
\label{Table2:BE2}
\end{table}

\clearpage

\begin{figure}
    \centering
    \includegraphics[width=0.75\textwidth]{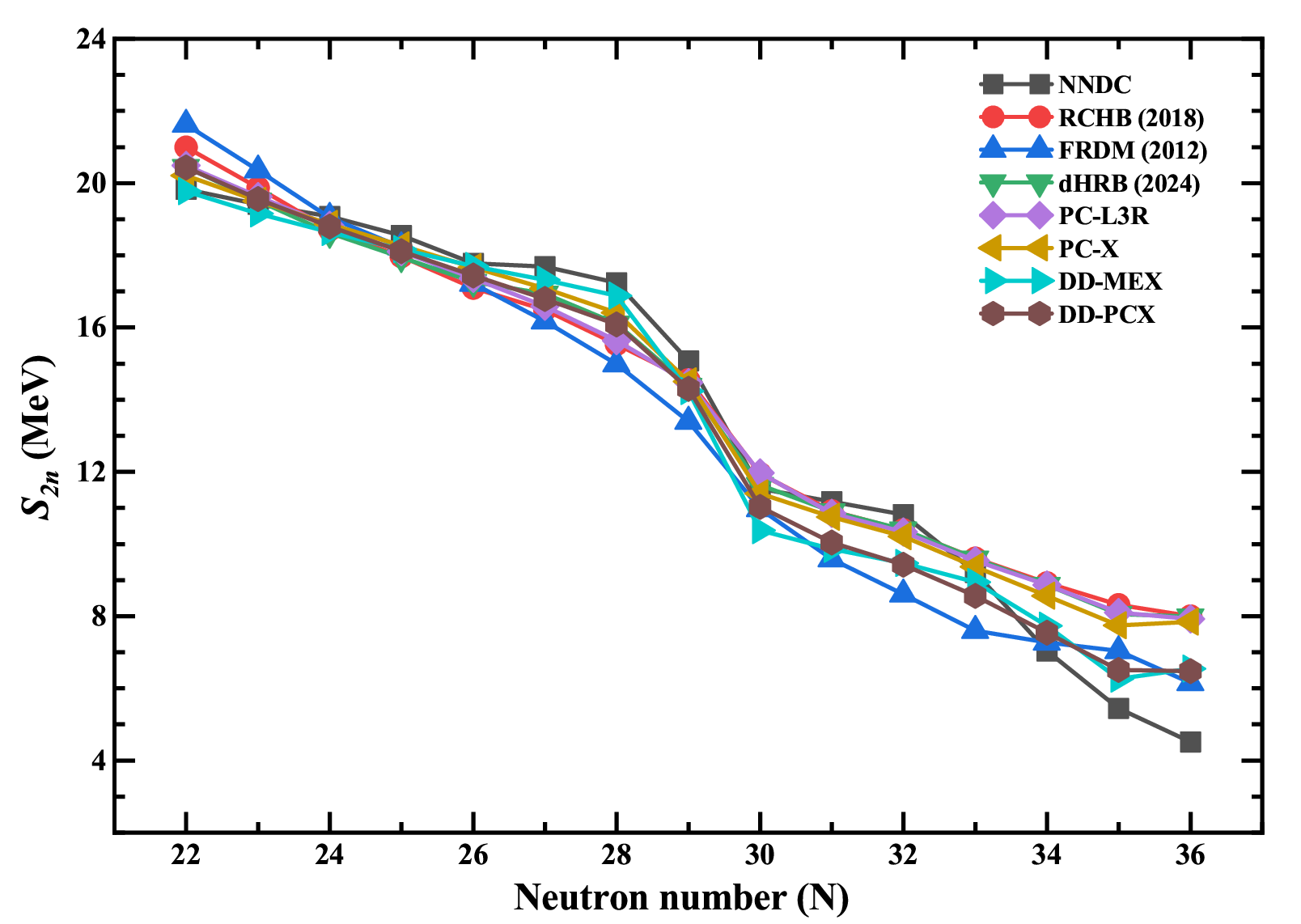}
    \caption{Two-neutron separation energies $S_{2n} (N, Z)$ of calcium isotopes (N=22-36) computed from Eq.~\ref{eq:S2n}, compared against experimental data from the NNDC database and predictions from seven theoretical frameworks: RCHB (2018), FRDM (2012), dHRB (2024), and the RHB parametrizations PC-L3R, PC-X, DD-MEX, and DD-PCX.}
    \label{fig:S2n}
\end{figure}

\begin{figure}
    \centering
    \includegraphics[width=0.75\textwidth]{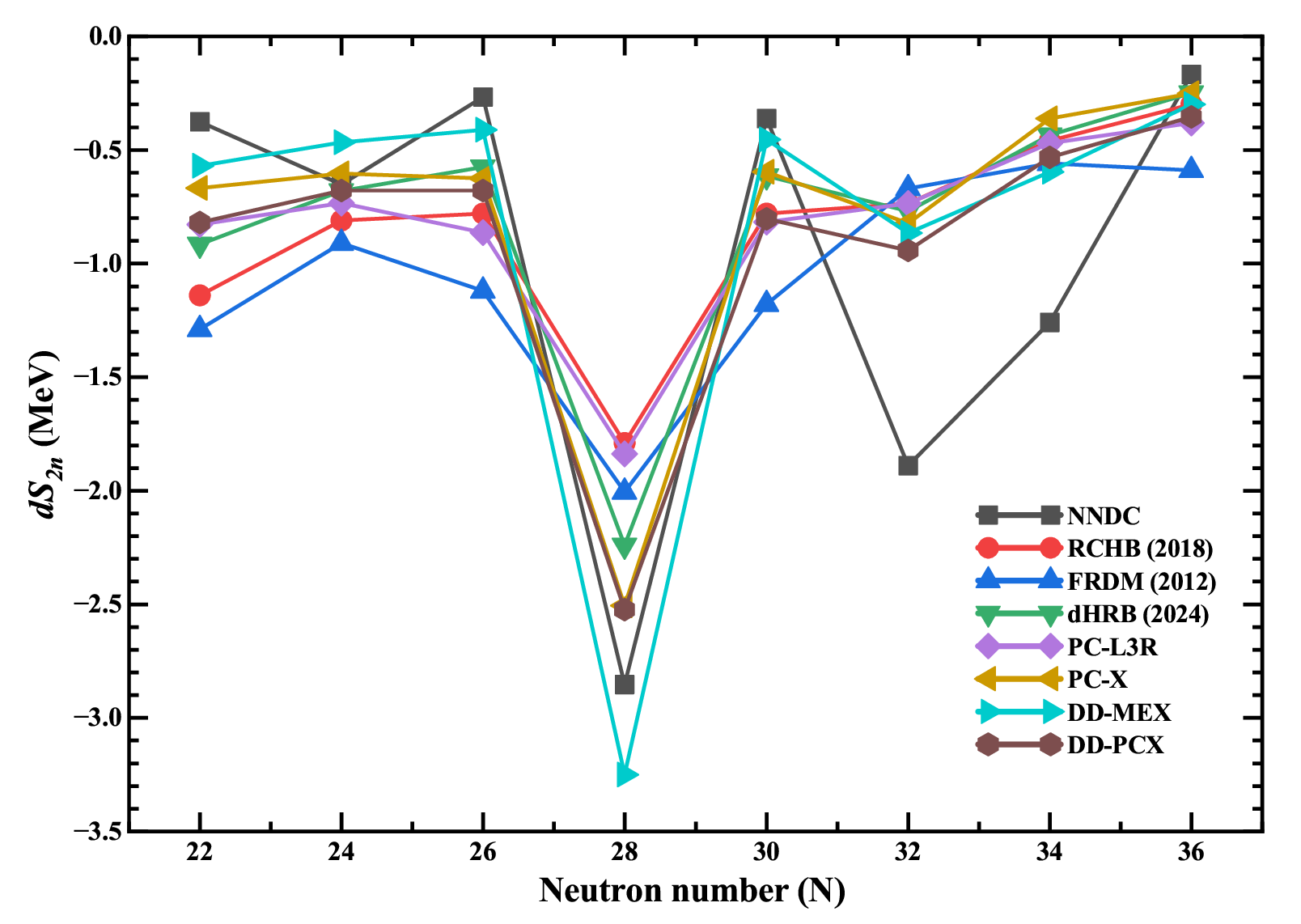}
    \caption{Differential two-neutron separation energies $dS_{2n}$ of calcium isotopes computed from Eq.~\ref{eq:dS2n}. Symbols and frameworks are the same as in Fig.\ref{fig:S2n}.}
    \label{fig:dS2n}
\end{figure}

\begin{figure}
    \centering
    \includegraphics[width=0.75\textwidth]{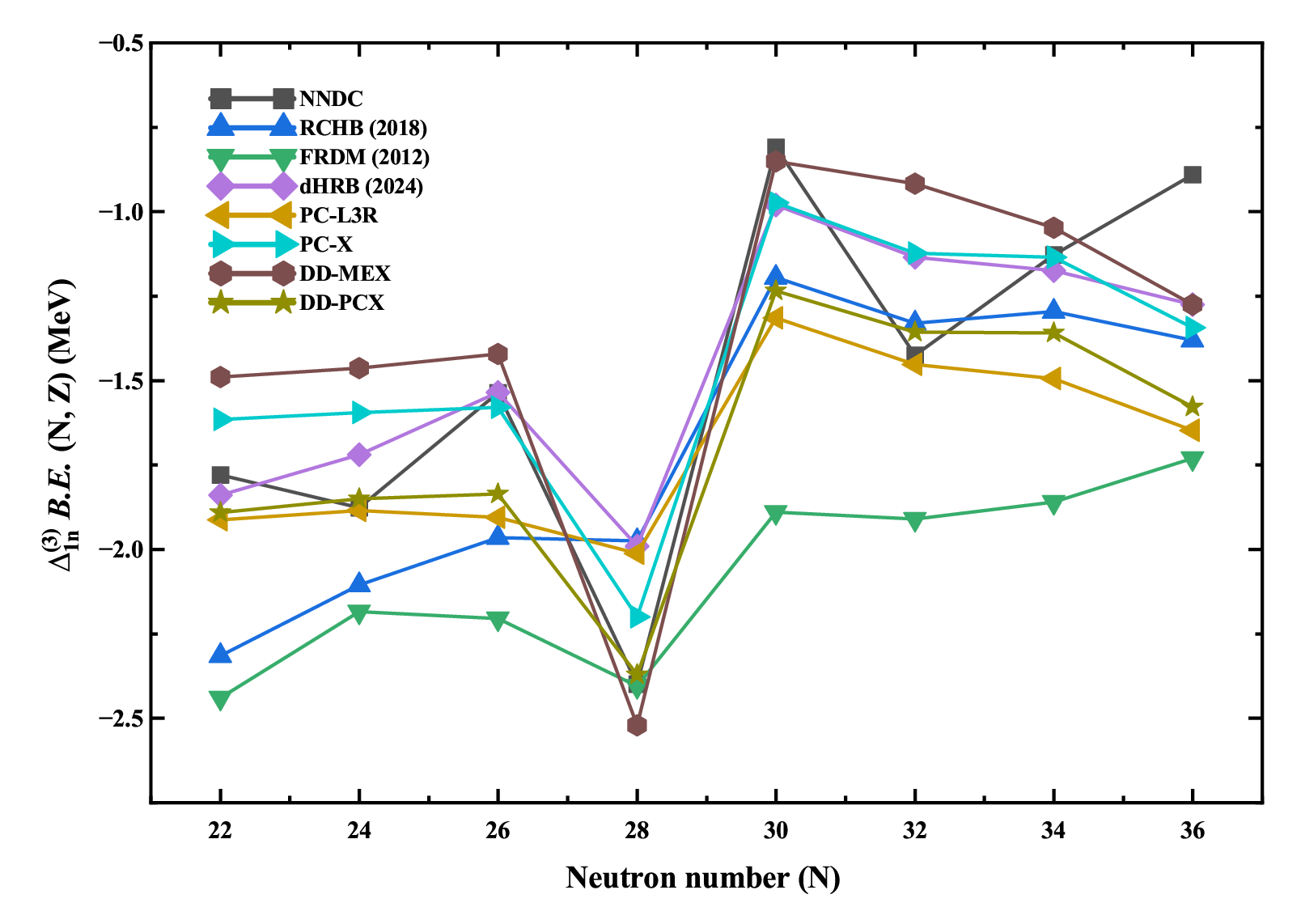}
    \caption{Three-point binding energy filter $\Delta_{1n}^{(3)} B.E. (N, Z)$ of calcium isotopes computed from Eq.~\ref{eq:three_point_mass_filter}. Symbols and frameworks are the same as in Fig.\ref{fig:S2n}.}
    \label{fig:delta1nBE}
\end{figure}

\begin{figure}
    \centering
    \includegraphics[width=0.75\textwidth]{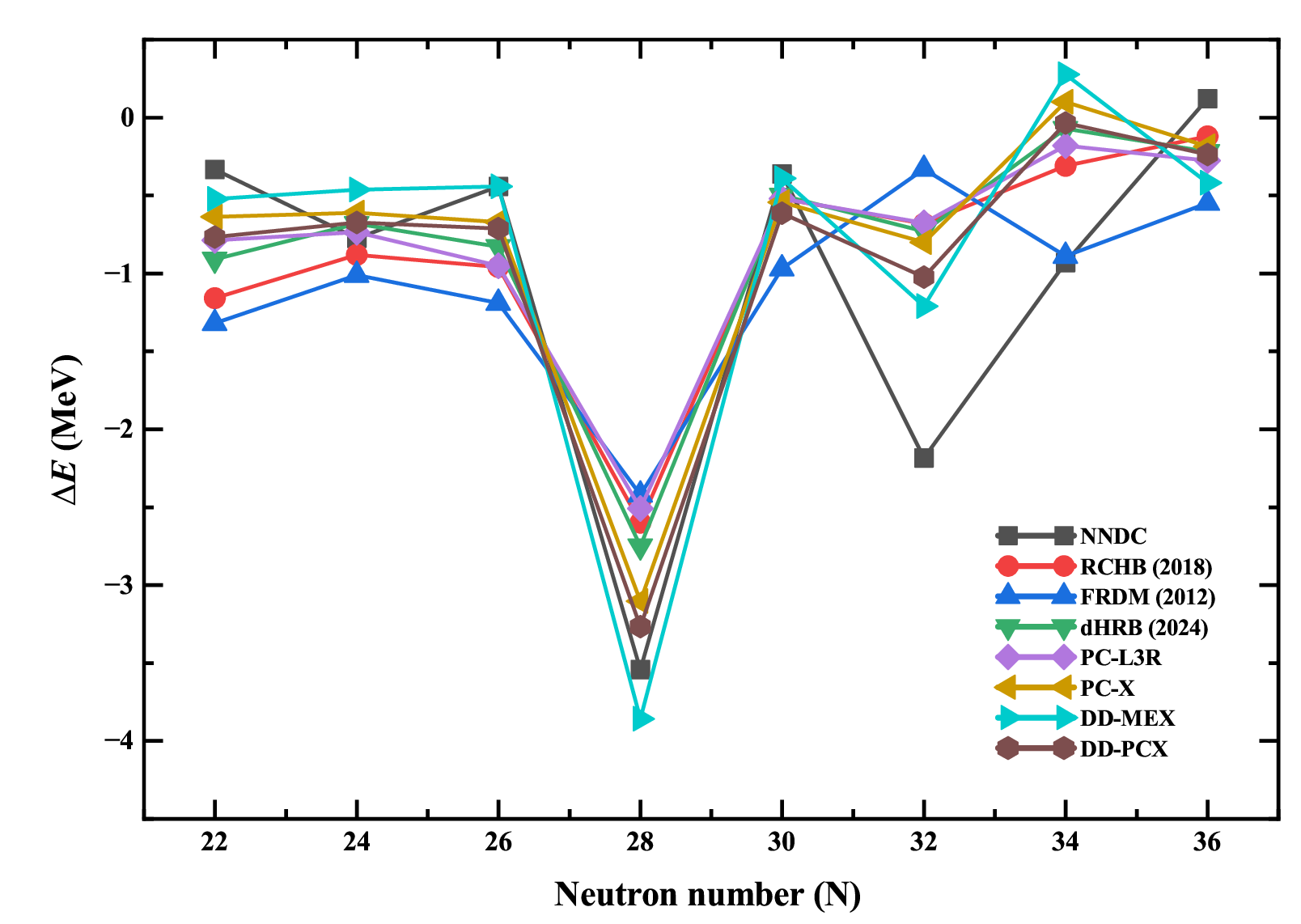}
    \caption{Neutron shell gap energies $\Delta E$ of even-mass calcium isotopes computed from Eq.~\ref{eq:neutron_shell_gap}. Symbols and frameworks are the same as in Fig.\ref{fig:S2n}.}
    \label{fig:deltaE}
\end{figure}

\begin{figure}
    \centering
    \includegraphics[width=0.75\textwidth]{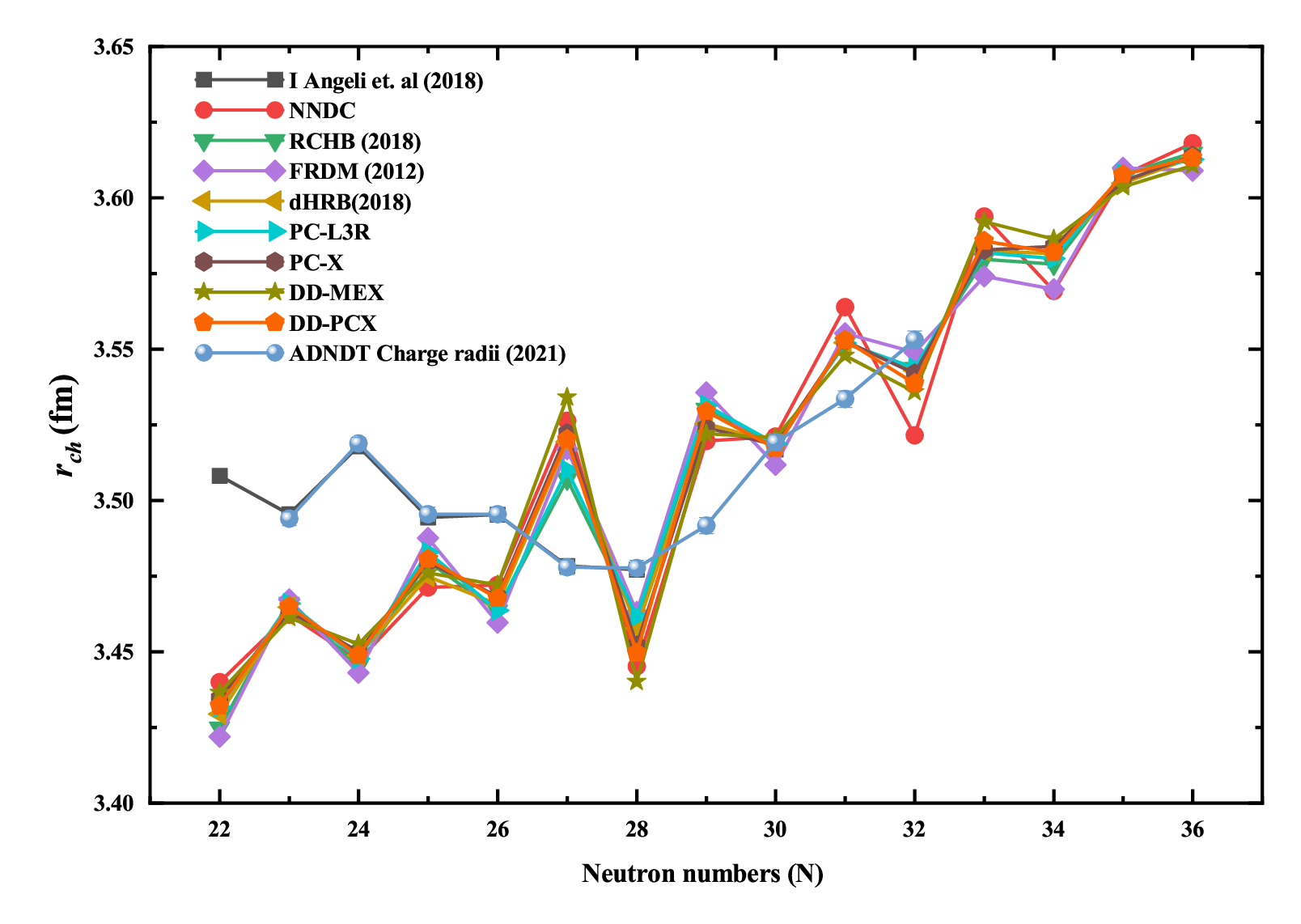}
    \caption{RMS charge radii as a function of neutron number for Calcium isotopes, compared against experimental data from I. Angeli et. al \cite{angeli2013table} and the ADNDT (2021) \cite{li2021compilation} compilation alongside predictions from the other theoretical frameworks listed in Fig.\ref{fig:S2n}.}
    \label{fig:r_ch}
\end{figure}

\begin{figure}
    \centering
    \includegraphics[width=0.75\textwidth]{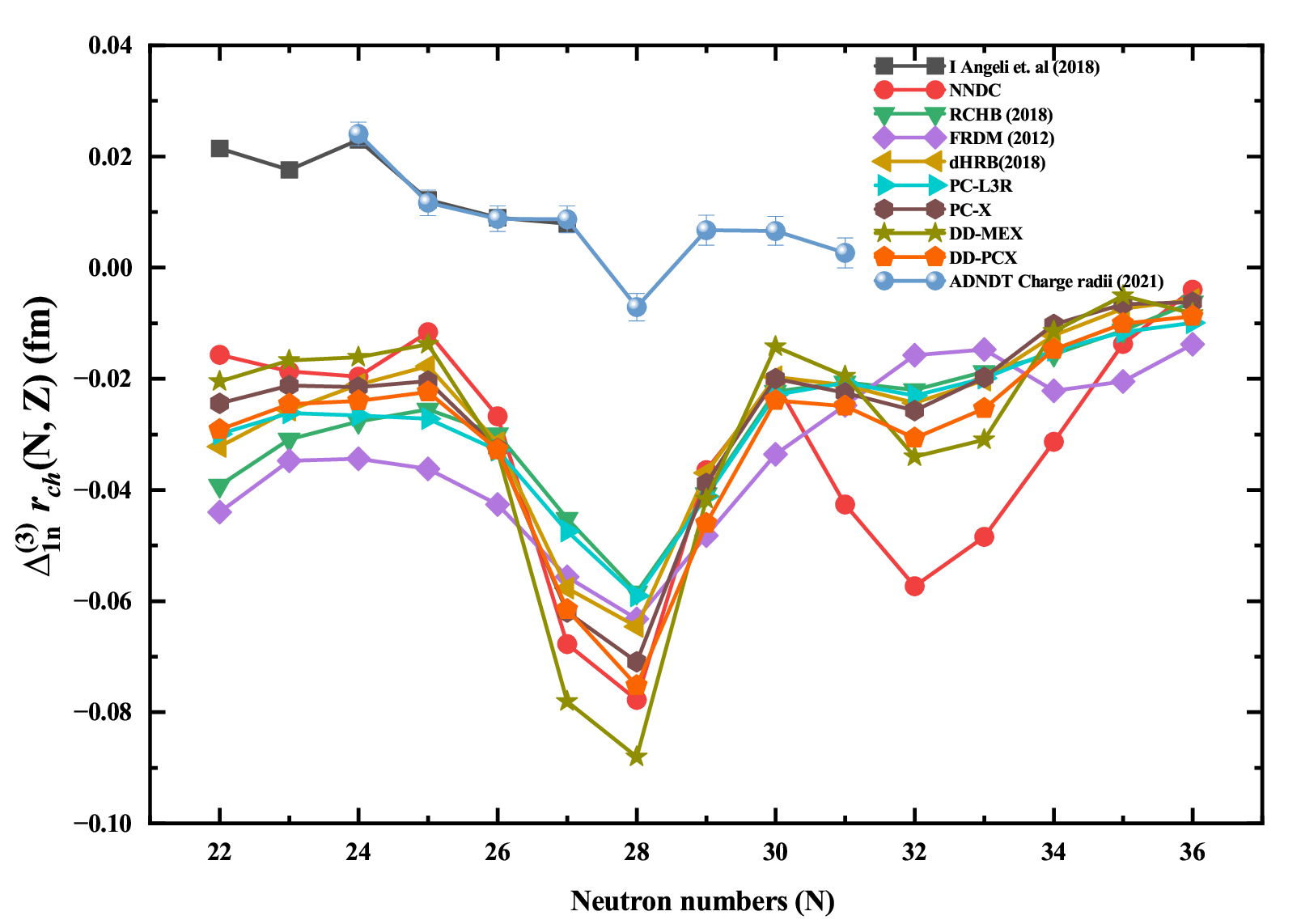}
    \caption{Three-point charge radii difference as a function of neutron number. Symbols and frameworks are the same as in Fig.\ref{fig:r_ch}.}
    \label{fig:delta1nRch}
\end{figure}

\end{document}